\documentclass{revtex4}
\input epsf
\usepackage{amsmath,amssymb}
\usepackage{graphicx}

\draft

\begin{document}
\titlepage
\title{Modified Friedmann Equations in R$^{-1}$-Modified Gravity}
\author{Xinhe Meng$^{1,2}$ \footnote{mengxh@public.tpt.tj.cn}
 \ \ Peng Wang$^1$ \footnote{pewang@eyou.com}
} \affiliation{1.  Department of Physics, Nankai University,
Tianjin 300071, P.R.China} \affiliation{2. Institute of
Theoretical Physics, CAS, Beijing 100080, P.R.China}

\begin{abstract}
Recently, corrections to Einstein-Hilbert action that become
important at small curvature are proposed. We discuss the first
order and second order approximations to the field equations
derived by the Palatini variational principle. We work out the
first and second order Modified Friedmann equations and present
the upper redshift bounds when these approximations are valid. We
show that the second order effects can be neglected on the
cosmological predictions involving only the Hubble parameter
itself, e.g. the various cosmological distances, but the second
order effects can not be neglected in the predictions involving
the derivatives of the Hubble parameter. Furthermore, the Modified
Friedmann equations fit the SNe Ia data at an acceptable level.

PACS numbers: 98.80.Bp, 98.65.Dx, 98.80.Es
\end{abstract}

\maketitle

\textbf{1. Introduction}

That our Universe expansion is currently in an accelerating phase
now seems well-established. The most direct evidence for this is
from the measurement of type Ia supernova \cite{Perlmutter}. Other
indirect evidences such as the observations of CMB by the WMAP
satellite \cite{Spergel}, 2dF and SDSS also seem supporting this.

But now the mechanism responsible for this acceleration is
unclear. Many authors introduce a mysterious cosmic fluid called
dark energy to explain this. There are now many possibilities of
the form of dark energy. The simplest possibility is a
cosmological constant arising from vacuum energy \cite{Peebles}.
Other possibilities include a dynamical scalar field called
quintessence \cite{Caldwell}, or an exotic perfect fluid called
Chaplygin gas \cite{Kamenshchik}.

On the other hand, some authors suggest that maybe there does not
exist such mysterious dark energy, but the observed cosmic
acceleration is a signal of our first real lack of understanding
of gravitational physics \cite{Lue}. An example is the braneworld
theory of Dvali et al. \cite{Dvali}. Recently, some authors
proposed to add a $R^{-1}$ term in the Einstein-Hilbert action to
modify the General Relativity (GR) \cite{Carroll, Capozziello}. By
varying with respect to the metric, this additional term will give
fourth order field equations. It was shown in their work that this
additional term can give accelerating solutions of the field
equations without the need of introducing dark energy. The
matching with observations is also discussed for $R^n$ model by
Capozziello \cite{Ca}.

Based on this modified action, Vollick \cite{Vollick} used
Palatini variational principle to derive the field equations. In
the Palatini formalism, instead of varying the action only with
respect to the metric, one views the metric and connection as
independent field variables and vary the action with respect to
them independently. This would give second order field equations.
In the original Einstein-Hilbert action, this approach gives the
same field equations as the metric variation. But now the field
equations are different from that of from the metric variation. In
ref.\cite{Dolgov}, Dolgov et al. argued that the fourth order
field equations following from the metric variation suffer serious
instability problem. If this is indeed the case, the Palatini
approach appears even more appealing, because the second order
field equations following from Palatini variation are free of this
sort of instability (see below for details). However, the most
convincing motivation to take the Palatini formalism seriously is
that the field equations following from it fit the SNe Ia data at
an acceptable level, see Sec.3.

However, the field equations following from the Palatini formalism
are too complicated to deal with directly. We have to use
perturbation expansion to get more amenable approximated
equations. In ref.\cite{Vollick}, this has been done up to first
order in the perturbation expansion. As will be shown in this
paper, perturbation approach is valid when redshift is smaller
than about 1. Thus when close to 1, how good are the first order
equations as an approximation to the full field equations? The
answer we get is rather interesting: the second order effects can
be neglected on the cosmological predictions involved only the
Hubble parameter itself, e.g. the various cosmological distances,
but the second order effect can not be neglected in the
predictions involved the derivatives of the Hubble parameter, e.g.
the deceleration parameter, the effective equation of state of the
Modified Friedmann equation, etc. Furthermore, the first order
equations fit the SNe Ia data at an acceptable level. Thus, we
have good reason to believe that the full field equations also fit
the SNe Ia data at an acceptable level.

\textbf{2. The Field Equations}

First, we briefly review the derivation of the field equations
using Palatini variational principle and the derivation of the
first order approximation to the field equations. For details, see
ref.\cite{Vollick}.

The field equations follow from the variation in Palatini approach
of the action
\begin{equation}
S=-\frac{1}{2\kappa}\int{d^4x\sqrt{-g}L(R)}+\int{d^4x\sqrt{-g}L_M}\label{action}
\end{equation}
where $\kappa =8\pi G$, $L$ is a function of the scalar curvature
$R$ and $L_M$ is the Lagrangian density for matter..

Varying with respect to $g_{\mu\nu}$ gives
\begin{equation}
L'(R)R_{\mu\nu}-\frac{1}{2}L(R)g_{\mu\nu}=-\kappa
T_{\mu\nu}\label{2.2}
\end{equation}
where $T_{\mu\nu}$ is the energy-momentum tensor given by
\begin{equation}
T_{\mu\nu}=-\frac{2}{\sqrt{-g}}\frac{\delta S_M}{\delta
g^{\mu\nu}}\label{2.3}
\end{equation}

In the Palatini formalism, the connection is not associated with
$g_{\mu\nu}$, but with $h_{\mu\nu}\equiv L'(R)g_{\mu\nu}$, which
is known from varying the action with respect to $\Gamma
^{\lambda}_{\mu\nu}$. Thus the Christoffel symbol with respect to
$h_{\mu\nu}$ is given by
\begin{equation}
\Gamma
^{\lambda}_{\mu\nu}=\{^{\lambda}_{\mu\nu}\}_g+\frac{1}{2L'}[2\delta
^{\lambda}_{(\mu}\partial
_{\nu)}L'-g_{\mu\nu}g^{\lambda\sigma}\partial
_{\sigma}L']\label{Christoffel}
\end{equation}
where the subscript $g$ signifies that this is the Christoffel
symbol with respect to the metric $g_{\mu\nu}$.

The Ricci curvature tensor and Ricci scalar is given by
\begin{equation}
R_{\mu\nu}=R_{\mu\nu}(g)-\frac{3}{2}(L')^{-2}\nabla _{\mu}L'\nabla
_{\nu}L'+(L')^{-1}\nabla _{\mu}\nabla
_{\nu}L'+\frac{1}{2}(L')^{-1}g_{\mu\nu}\nabla _{\sigma}\nabla
^{\sigma}L'\label{Ricci}
\end{equation}
\begin{equation}
R=R(g)+3(L')^{-1}\nabla _{\mu}\nabla ^{\mu}
L'-\frac{3}{2}(L')^{-2}\nabla_{\mu}L'\nabla^{\mu}L'\label{scalar}
\end{equation}
where $R_{\mu\nu}(g)$ is the Ricci tensor with respect to
$g_{\mu\nu}$ and $R=g^{\mu\nu}R_{\mu\nu}$. Note by contracting
eq.(\ref{2.2}), we can solve $R$ as a function of $T$:
\begin{equation}
L'(R)R-2L(R)=-\kappa T\label{R(T)}
\end{equation}
Thus (\ref{Ricci}), (\ref{scalar}) do define the Ricci tensor with
respect to $h_{\mu\nu}$.

Now apply the above Palatini formalism to the special $L(R)$
suggested in ref.\cite{Carroll, Capozziello}:
\begin{equation}
L(R)=R-\frac{\alpha ^2}{3R}\label{L(R)}
\end{equation}
where $\alpha$ is a positive constant with the same dimensions as
$R$ and following \cite{Vollick}, the factor of 3 is introduced to
simplify the field equations.

The field equations follow from eq.(\ref{2.2})
\begin{equation}
(1+\frac{\alpha
^2}{3R^2})R_{\mu\nu}-\frac{1}{2}g_{\mu\nu}(R-\frac{\alpha
^2}{3R})=-\kappa T_{\mu\nu}\label{field equ}
\end{equation}

Contracting the indices gives
\begin{equation}
R=\frac{1}{2}\alpha [\frac{\kappa
T}{\alpha}\pm2\sqrt{1+\frac{1}{4}(\frac{\kappa
T}{\alpha})^2}]\label{R}
\end{equation}
Since $T$ is negative and for large $|T|$ we expect the above to
reduce to $R=\kappa T$, we just take the minus sign in the
following discussions.

From eq.(\ref{field equ}) and eq.(\ref{R}) we can see that the
field equations reduce to the Einstein equations if $|\kappa T|
\gg \alpha$. On the other hand, when $\alpha \gg |\kappa T|$,
deviations from the Einstein's theory will be large. This is
exactly the case we are interested in. We hope it can explain
today's cosmic acceleration.

Recently, Dolgov \cite{Dolgov} argued that for a given
$T_{\mu\nu}$, the field equation for Ricci scalar derived by
metric variation suffers serious instability problem. Now in the
case of Palatini formalism, it can be seen from equations
(\ref{R(T)}) and (\ref{scalar}) that, for a given $T_{\mu\nu}$, we
can directly get $R(g)$ without any need to solve differential
equations, see eq.(\ref{R}). Thus there is no instabilities of
this sort. This makes the Palatini formalism more appealing and
worth further investigations.

Now consider the Robertson-Walker metric describing the
cosmological evolution,
\begin{equation}
ds^2=-dt^2+a(t)^2(dx^2+dy^2+dz^2)\label{metric}
\end{equation}
We consider only the spatially flat case, i.e. $k=0$, which is now
favored by CMB observations \cite{Spergel} and also is the
prediction of inflation theory \cite{Liddle}.

We assume $\alpha \gg |\kappa T|$. In ref.\cite{Vollick}, it was
shown that, up to first order in $\kappa T/\alpha$, the scalar
curvature can be obtained from eq.(\ref{R})
\begin{equation}
R\simeq \alpha (-1+\frac{\kappa T}{2\alpha})\label{1orderR}
\end{equation}

Then from eq.(\ref{L(R)}),
\begin{equation}
L\simeq -\frac{2}{3}\alpha (1-\frac{\kappa T}{\alpha})\label{1stL}
\end{equation}
\begin{equation}
L'\simeq \frac{4}{3}(1+\frac{\kappa T}{4\alpha}) \label{1stL'}
\end{equation}

The first order field equation follows from eq.(\ref{field equ})
\begin{equation}
R_{\mu\nu}=-\frac{1}{4}\alpha g_{\mu\nu}-\kappa
(\frac{3}{4}T_{\mu\nu}-\frac{5}{16}Tg_{\mu\nu}) \label{1stfield}
\end{equation}

The non-vanishing components of the Ricci tensor up to first order
follow from eq.(\ref{Ricci}) and eq.(\ref{metric})
\begin{equation}
R_{00}=3\frac{\ddot{a}}{a}+\frac{3\kappa
\ddot{T}}{8\alpha}\label{1stR00}
\end{equation}
\begin{equation}
R_{ij}=-(a\ddot{a}+2\dot{a}^2+\frac{a\dot{a}}{4}\frac{\kappa
\dot{T}}{\alpha}+\frac{a^2}{8}\frac{\kappa
\ddot{T}}{\alpha})\delta _{ij}\label{1stRij}
\end{equation}

Assume the matter in the recent cosmological times contains only
dust with $T=-\rho_0/a^3$, where $\rho_0$ is the present energy
density of dust. Then
\begin{equation}
\dot{T}=3\rho \frac{\dot{a}}{a},\  \ \ddot{T}=3\rho
(\frac{\ddot{a}}{a}-4(\frac{\dot{a}}{a})^2)\label{T'}
\end{equation}

Now the first order equation describing the evolution of Hubble
parameter $H\equiv \dot{a}/a$ in this modified gravity theory,
i.e. the Modified Friedmann (MF) equation, can be get from
equations (\ref{1stR00}), (\ref{1stRij}), (\ref{T'}) and
(\ref{1stfield}):
\begin{equation}
H^2=\frac{\frac{11}{8}\kappa
\rho+\frac{\alpha}{2}}{6+\frac{9}{4}\frac{\kappa
\rho}{\alpha}}\label{1stMF}
\end{equation}
It is interesting to note that the coefficient $\alpha$ of the
$R^{-1}$ term in the modified Einstein-Hilbert action just behaves
like a cosmological constant in the recent cosmological times. In
ref.\cite{Carroll}, it has already been shown that its magnitude
should be about $10^{-67} (eV)^2$ in order to be consistent with
today's comic acceleration. This is now easy to interpret.

Now we must check whether the assumption $\alpha \gg |\kappa T|$
holds and when this assumption breaks down. In order to do this,
define $z_E$ by $\alpha=\kappa\rho_0(1+z_E)^3$. This parameter
gives the upper redshift beyond which the assumption breaks down.
In terms of $z_E$, eq.(\ref{1stMF}) can be rewritten as
\begin{equation}
H^2=H_0^2\Omega_m\frac{\frac{11}{8}(1+z)^3+\frac{1}{2}(1+z_E)^3}{2+\frac{3}{4}(\frac{1+z}{1+z_E})^3}\label{1stMF2}
\end{equation}
where as usual, $\Omega_m=\rho_0/\rho_{c0}$. $\rho_0$, $H_0$ is
the present matter density and Hubble parameter respectively and
$\rho_{c0}=3H_0^2/\kappa$. In the following discussions and
numerical computations, we always take the value
$H_0=70kmMpc^{-1}sec^{-1}$ which is now favored by CMB
observations \cite{Spergel}.

Setting $z=0$ in eq.(\ref{1stMF2}) gives
\begin{equation}
z_E=(\frac{\frac{11}{8}\Omega_m-2+\sqrt{(\frac{11}{8}\Omega_m-2)^2+\frac{3}{2}\Omega_m}}{3/2})^{-\frac{1}{3}}-1
\label{1stzE}
\end{equation}

\begin{figure}
  \includegraphics[width=0.8\columnwidth]{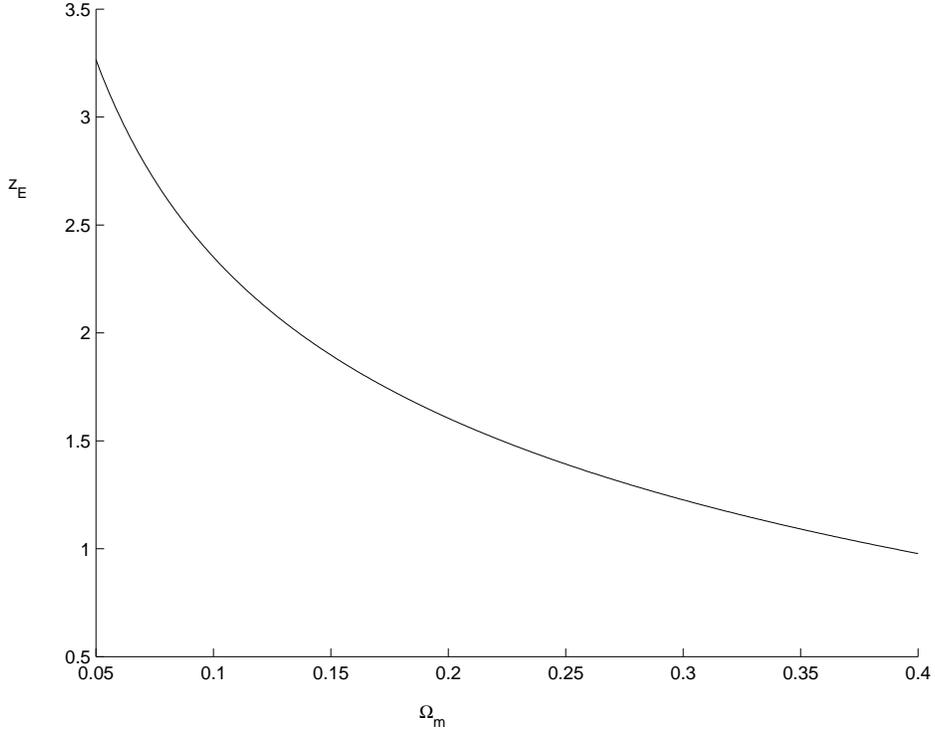}
  \caption{The dependence of the critical redshift $z_E$ which characterizes the valid region of
  perturbation expansion of the field equations on the matter energy density fraction $\Omega_m$}\label{1}
\end{figure}

The dependence of $z_E$ on $\Omega_m$ is drawn in Fig.1. In
particular, $\Omega_m=0.1, 0.2, 0.3$ gives $z_E=2.351, 1.603,
1.226$ respectively. We will use these values in the following
discussions. Thus, the first order Modified Friedmann equation is
at most a good approximation to the full field equation up to,
e.g. in the case of $\Omega_m=0.3$, about redshift 1. We want to
know up to what redshift first order approximation is good enough.
This is not so obvious, because when $z\sim 1$, $\kappa
T/\alpha\sim 0.73$. This is not very small, thus the effects of
higher order terms in perturbation expansion may not be negligible
when apply the MF equation on these redshift close to $z_E$.
Future high redshift supernova observations would reach such high
redshift region so the behavior of the MF equation in these region
is important. On the other hand, if we find that the first order
equation is doing good in describing our Universe's evolution up
to redshift $z_E$ and the second order correction is small, this
is a strong indication that the full field equations also behave
well in describing the cosmological evolution. In the below we
will see that the effects of the second correction is rather
interesting: the first order equation fits the SNe Ia data in an
acceptable level and the second order correction is negligible up
to the critical redshift $z_E$. But the second order effects can
not be neglected in describing the acceleration rate, i.e. the
deceleration parameter. Thus we have good reason to believe the
full field equation is also good in fitting the supernova data and
this modified gravity theory may be a good candidate for
explanation of the cosmic acceleration.

Thus in order to investigate the valid region of the first order
approximation, let us consider the second order approximation to
the theory.

First, the scalar curvature can be obtained from eq.(\ref{R})
\begin{equation}
R\simeq \alpha[-1+\frac{\kappa
T}{2\alpha}-\frac{1}{8}(\frac{\kappa T}{\alpha})^2]\label{2ndR}
\end{equation}

From eq.(\ref{L(R)}):
\begin{equation}
L\simeq \alpha[-\frac{2}{3}+\frac{2}{3}\frac{\kappa
T}{\alpha}-\frac{1}{12}(\frac{\kappa T}{\alpha})^2]\label{2ndL}
\end{equation}
\begin{equation}
L'\simeq \frac{4}{3}[1+\frac{\kappa
T}{4\alpha}+\frac{1}{8}(\frac{\kappa T}{\alpha})^2]\label{2ndL'}
\end{equation}

The second order field equation can be obtained from
eq.(\ref{field equ})
\begin{equation}
R_{\mu\nu}=-\frac{1}{4}\alpha g_{\mu\nu}-\kappa
[\frac{3}{4}T_{\mu\nu}-\frac{5}{16}Tg_{\mu\nu}]+\frac{\kappa^2T}{4\alpha}[\frac{3}{4}T_{\mu\nu}-\frac{5}{16}
Tg_{\mu\nu}]\label{2ndfield}
\end{equation}

The non-vanishing component of Ricci curvature tensor up to second
order is obtained from eq.(\ref{Ricci})
\begin{equation}
R_{00}=3\frac{\ddot{a}}{a}+\frac{3}{8}\frac{\kappa
\ddot{T}}{\alpha}+\frac{15}{64}(\frac{\kappa
\dot{T}}{\alpha})^2+\frac{9}{32}\frac{\kappa
T}{\alpha}\frac{\kappa \ddot{T}}{\alpha}\label{2ndR00}
\end{equation}
\begin{equation}
R_{ij}=-[a\ddot{a}+2\dot{a}^2+\frac{a\dot{a}}{4}\frac{\kappa
\dot{T}}{\alpha}+\frac{a^2}{8}\frac{\kappa
\ddot{T}}{\alpha}+\frac{9a^2}{64}(\frac{\kappa
\dot{T}}{\alpha})^2+\frac{3a\dot{a}}{16}\frac{\kappa
T}{\alpha}\frac{\kappa
\dot{T}}{\alpha}+\frac{3a^2}{32}\frac{\kappa
T}{\alpha}\frac{\kappa \ddot{T}}{\alpha}]\delta_{ij}\label{2ndRij}
\end{equation}

The second order Modified Friedmann equation now follows from
equations (\ref{2ndR00}), (\ref{2ndRij}) and (\ref{2ndfield}):
\begin{equation}
H^2=\frac{\frac{11}{8}\kappa
\rho+\frac{11}{32}\frac{(\kappa\rho)^2}{\alpha}+\frac{\alpha}{2}}{6+\frac{9}{4}\frac{\kappa
\rho}{\alpha}}\label{2ndMF}
\end{equation}
The only correction to the first order equation is the quadratic
term in the numerator. It is interesting to note that while the
first order MF equation is formally similar to the Friedmann
equation in the $\Lambda$CDM model, the second order MF equation
is formally similar to the modified Friedmann equation in the RS
II brane world cosmology model with an effective cosmological
constant \cite{Wang}.

In terms of the $z_E$ defined above, eq.(\ref{2ndMF}) can be
rewritten as
\begin{equation}
H^2=H_0^2\Omega_m\frac{\frac{11}{8}(1+z)^3+\frac{11}{32}\frac{(1+z)^6}{(1+z_E)^3}+\frac{1}{2}(1+z_E)^3}{
2+\frac{3}{4}(\frac{1+z}{1+z_E})^3}\label{2ndMF2}
\end{equation}

Set $z=0$ in eq.(\ref{2ndMF2}) gives
\begin{equation}
z_E=(\frac{\frac{11}{8}\Omega_m-2+\sqrt{(\frac{11}{8}\Omega_m-2)^2+(\frac{3}{2}-\frac{11}{16}\Omega_m)
\Omega_m}}{3/2-\frac{11}{16}\Omega_m})^{-\frac{1}{3}}-1\label{2ndze}
\end{equation}

The dependence of $z_E$ on $\Omega_m$ is almost identical to the
first order equation. The difference is very small and is of about
order $10^{-4}$.

\textbf{3. Data Fitting with SNe Ia Observations}

It is the observations of the SNe Ia that first reveal our
Universe is in an accelerating phase. It is still the most
important evidence for acceleration and the best discriminator
between different models to explain the acceleration. Thus, any
model attempting to explain the acceleration should fit the SNe Ia
data as the basic requirement.

\begin{figure}
  \includegraphics[width=0.8\columnwidth]{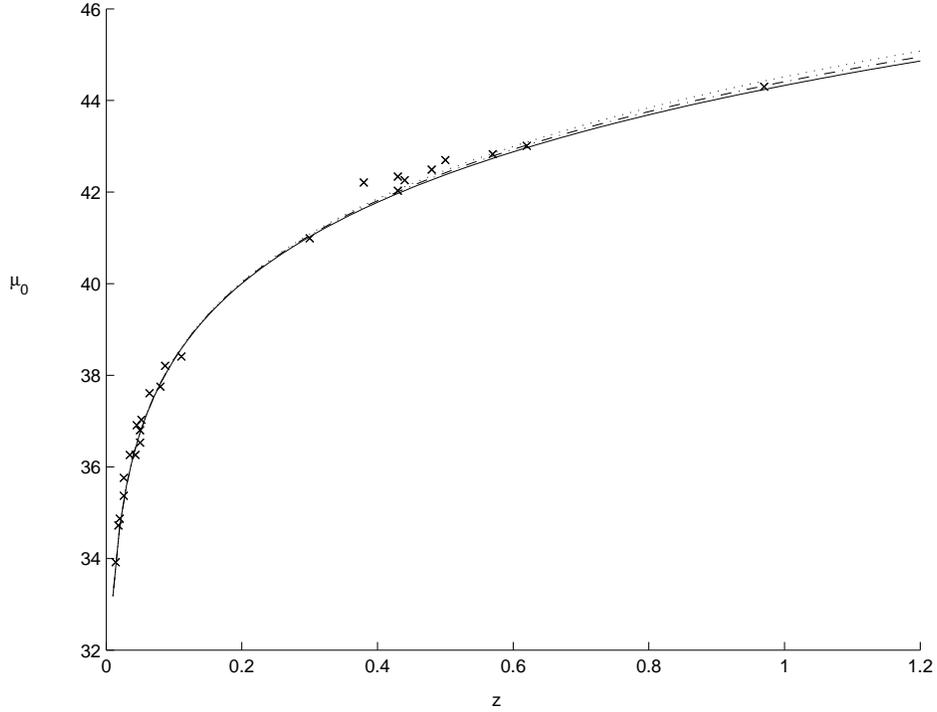}
  \caption{The dependence of luminosity distance on redshift computed from the first order modified Friedmann
  equation. The dotted, dashed and solid lines correspond to $\Omega_m=0.1, 0.2, 0.3$, respectively. The little crosses
  are the observed data }\label{1}
\end{figure}

\begin{figure}
  \includegraphics[width=0.8\columnwidth]{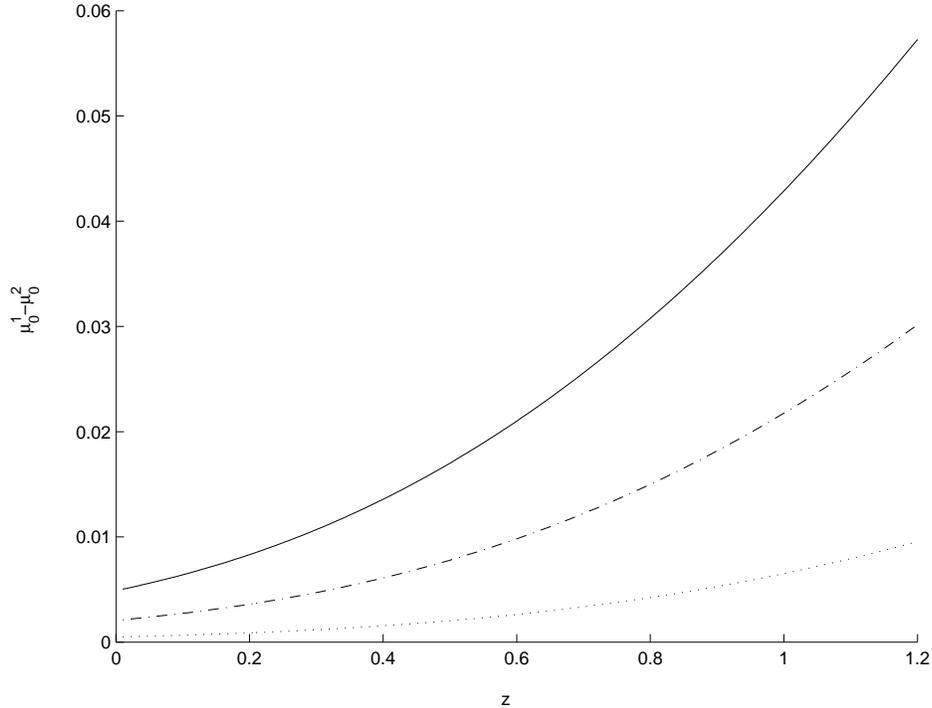}
  \caption{The difference between the luminosity distances computed from the first order and second order
  modified Friedmann equations. The dotted, dashed and solid lines correspond to $\Omega_m=0.1, 0.2, 0.3$, respectively. The difference is very small even up to high redshift larger than 1 which is
  the upper limit for the validity of perturbation expansion of the full field equation. It shows that we can
  trust the prediction of first order equation in calculating luminosity distance in almost all the redshift region smaller than the critical value
  $z_E$.  }\label{1}
\end{figure}

\begin{figure}
  \includegraphics[width=0.8\columnwidth]{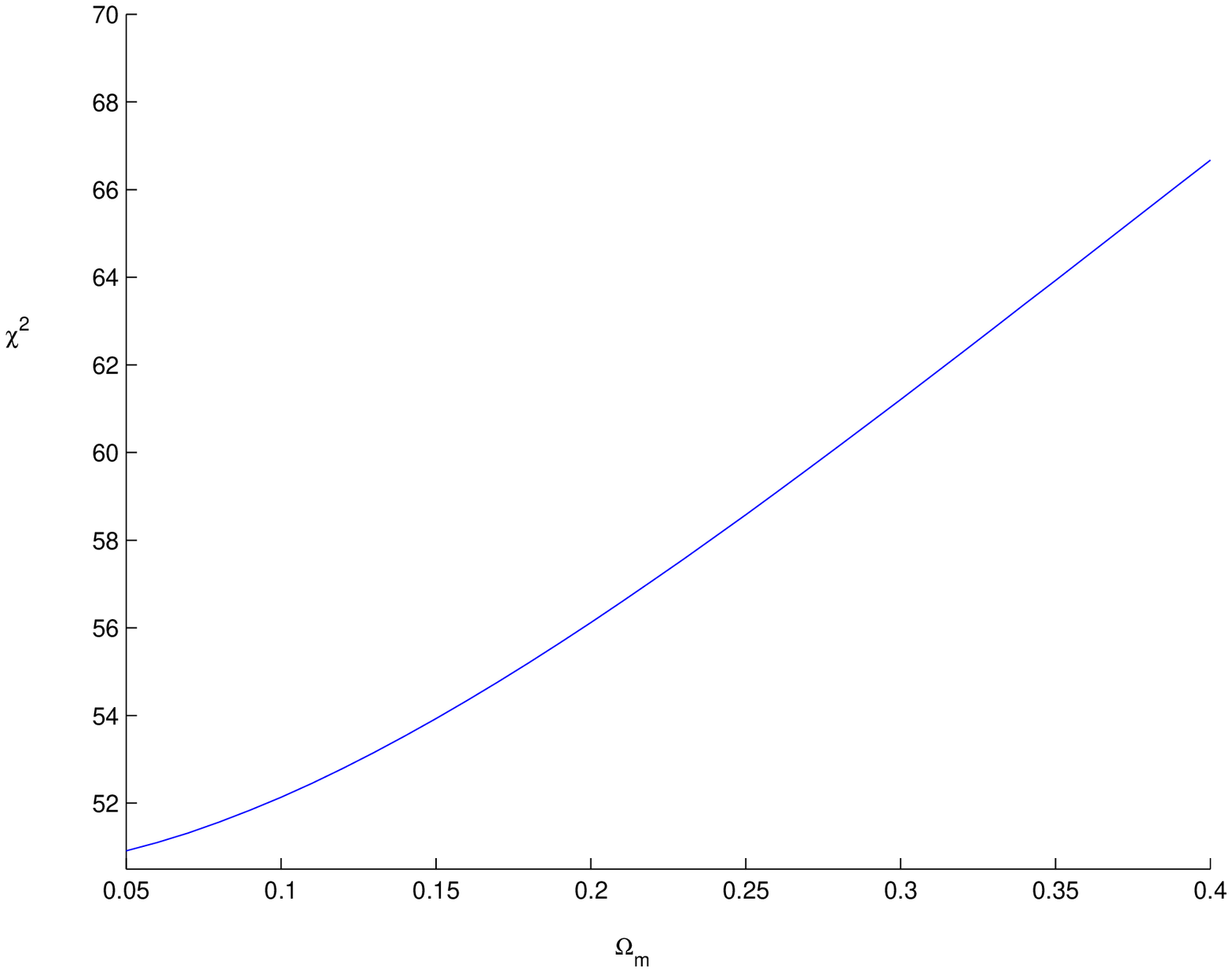}
  \caption{The dependence of the $\chi^2$ on the parameter $\Omega_m$. It can be seen that $\chi^2$ gets a little smaller for
smaller value of $\Omega_m$.}\label{1}
\end{figure}

The supernova observation is essentially a determination of
redshift-luminosity distance relationship. The luminosity distance
is obtained by a standard procedure:
\begin{equation}
D_L=(1+z)r=(1+z)\int_{0}^{z}\frac{dz'}{H(z')}\label{7}
\end{equation}
where $H(z)$ is given by eq.(\ref{1stMF2}) or eq.(\ref{2ndMF2})
for first order and second order approximation respectively.

In the data fitting, we actually compute the quantity,
\begin{equation}
\mu _0=5\log(\frac{D_L}{Mpc})+25\label{8}
\end{equation}

The quality of the fitting is characterized by the parameter:
\begin{equation}
\chi ^2=\sum_i \frac{[\mu _{0,i}^o-\mu _{0,i}^t]^2}{\sigma _{\mu
 _0,i}^2+\sigma^2_{mz,i}}\label{chi}
\end{equation}
where $\mu _{0,i}^o$ is the observed value, $\mu _{0,i}^t$ is the
value calculated through the model described above, $\sigma
^2_{\mu _0,i}$ is the measurement error, $\sigma^2_{mz,i}$ is
dispersion in the distance modulus due to the dispersion in galaxy
redshift caused by peculiar velocities. This quantity will be
taken as
\begin{equation}
\sigma_{mz}=\frac{\partial \log D_L}{\partial z}\sigma_z\label{}
\end{equation}
where following ref.\cite{Perlmutter}, $\sigma_z=200 km/s$. We use
data listed in ref.\cite{Fabris}, which contains 25 SNe Ia
observations. Since our purpose is to show that first order
approximation fits the data at an acceptable level and second
order effects can be ignored even up to high redshift such as 1,
also because trustable observations around redshift 1 is very few,
we think 25 low redshift samples is enough. Also we do not perform
a detailed $\chi ^2$ analysis, this is suitable when we get more
high redshift samples.

Fig.2 shows the prediction of the first order MF equation for
$\Omega_m=0.1, 0.2, 0.3$ respectively.

Fig.3 draws the difference between the $\mu_0$ computed using
first and second order MF equations for $\Omega_m=0.1, 0.2, 0.3$
respectively. The prediction of second order MF equation is almost
indistinguishable from the first order equation up to redshift 1.
The difference is about the order $10^{-2}$ even up to redshift 1.
So we did not draw the corresponding luminosity distance curves
for second order approximation. And this confirms the assertion we
made above. We do can trust that the predictions made by the first
order equation in calculating the luminosity distances are  good
approximation  to the full field equations even up to high
redshift, just not exceeding the critical redshift $z_E$.

Fig.4 shows the dependence of the $\chi^2$ computed using
eq.(\ref{chi}) on the $\Omega_m$. We can see that $\chi^2$ gets a
little smaller for smaller value of $\Omega_m$. We think this is
an interesting feature of this modified gravity theory: It has
already provide the possibility of eliminating the necessity for
dark energy for explanation of cosmic acceleration, now it can be
consistent with the SNe Ia observations without the assumption of
dark matter. Thus we boldly suggest that maybe this modified
gravity theory can provide the possibility of eliminating dark
matter (There have been some efforts in this direction, see
ref.\cite{Milgrom}). This is surely an interesting thing and worth
some investigations.

However, we should also note that we only use SNe Ia data smaller
than redshift 1 to draw the above conclusions. It can easily been
seen in Fig.2 that the differences between predictions drawn from
the different values of $\Omega_m$ become larger when redshift is
around 1. Thus, future high redshift supernova observations may
give a more conclusive discrimination between the parameters.

\textbf{4. The Deceleration Parameter}

\begin{figure}
  \includegraphics[width=0.4\columnwidth]{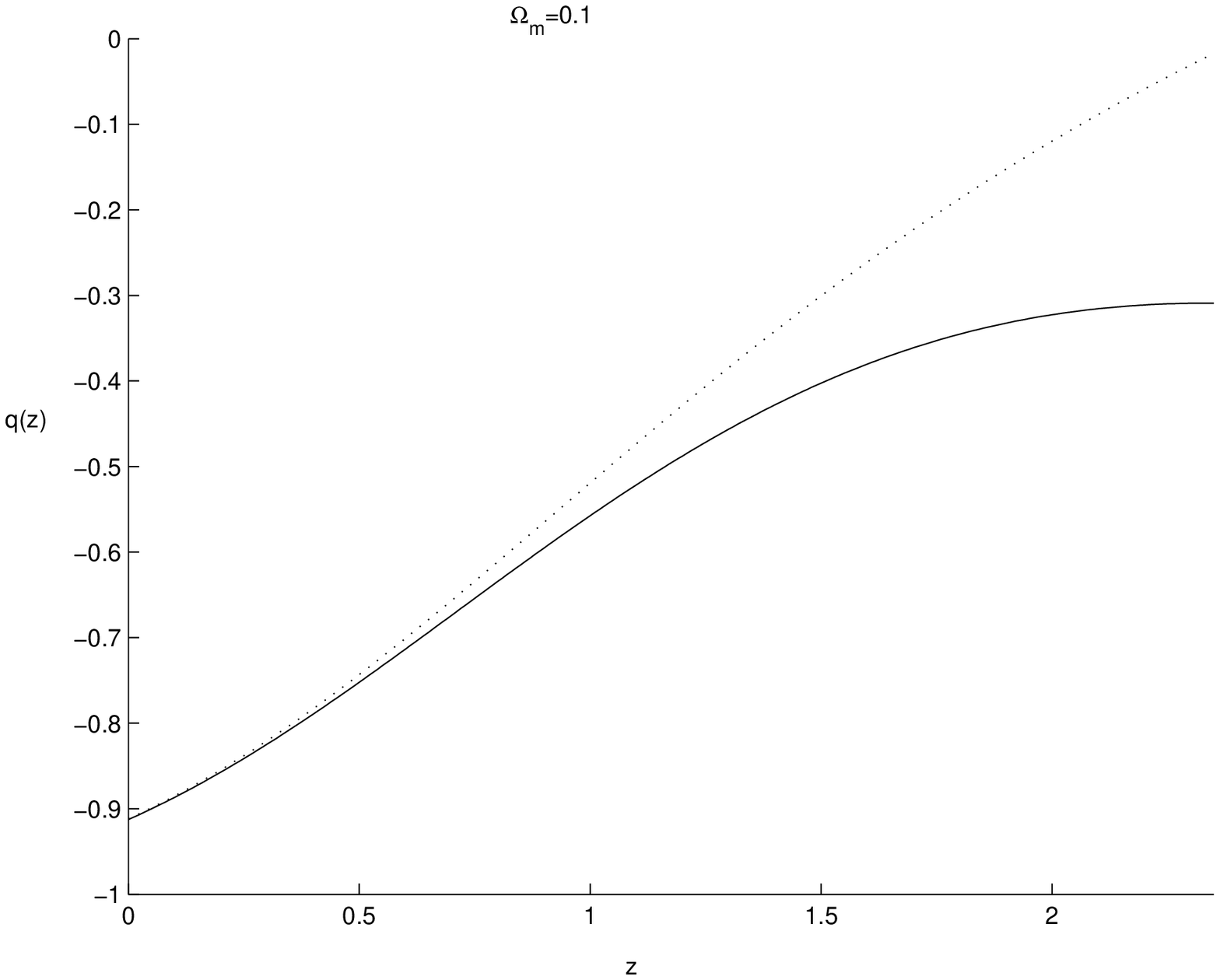}
  \includegraphics[width=0.4\columnwidth]{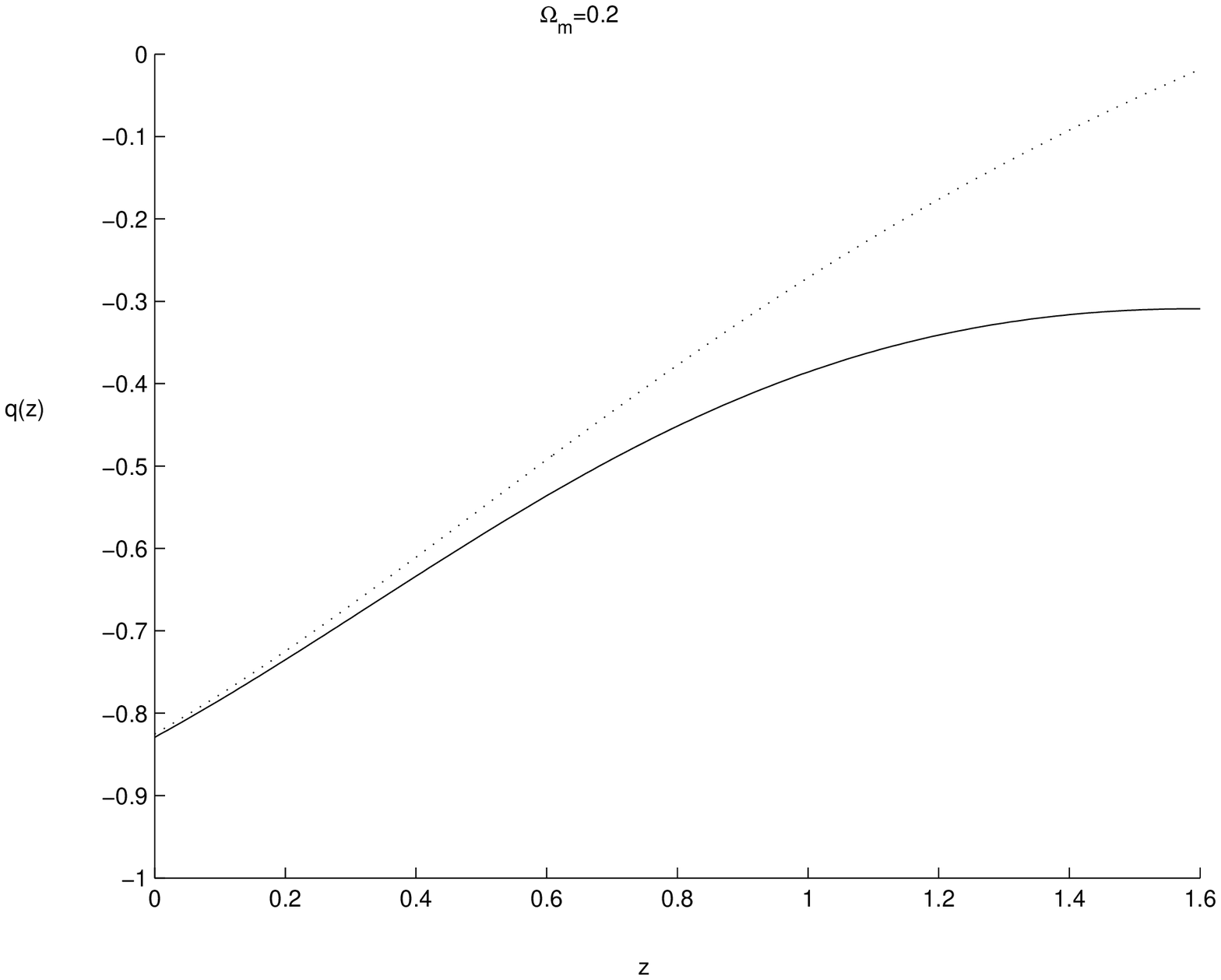}
  \includegraphics[width=0.4\columnwidth]{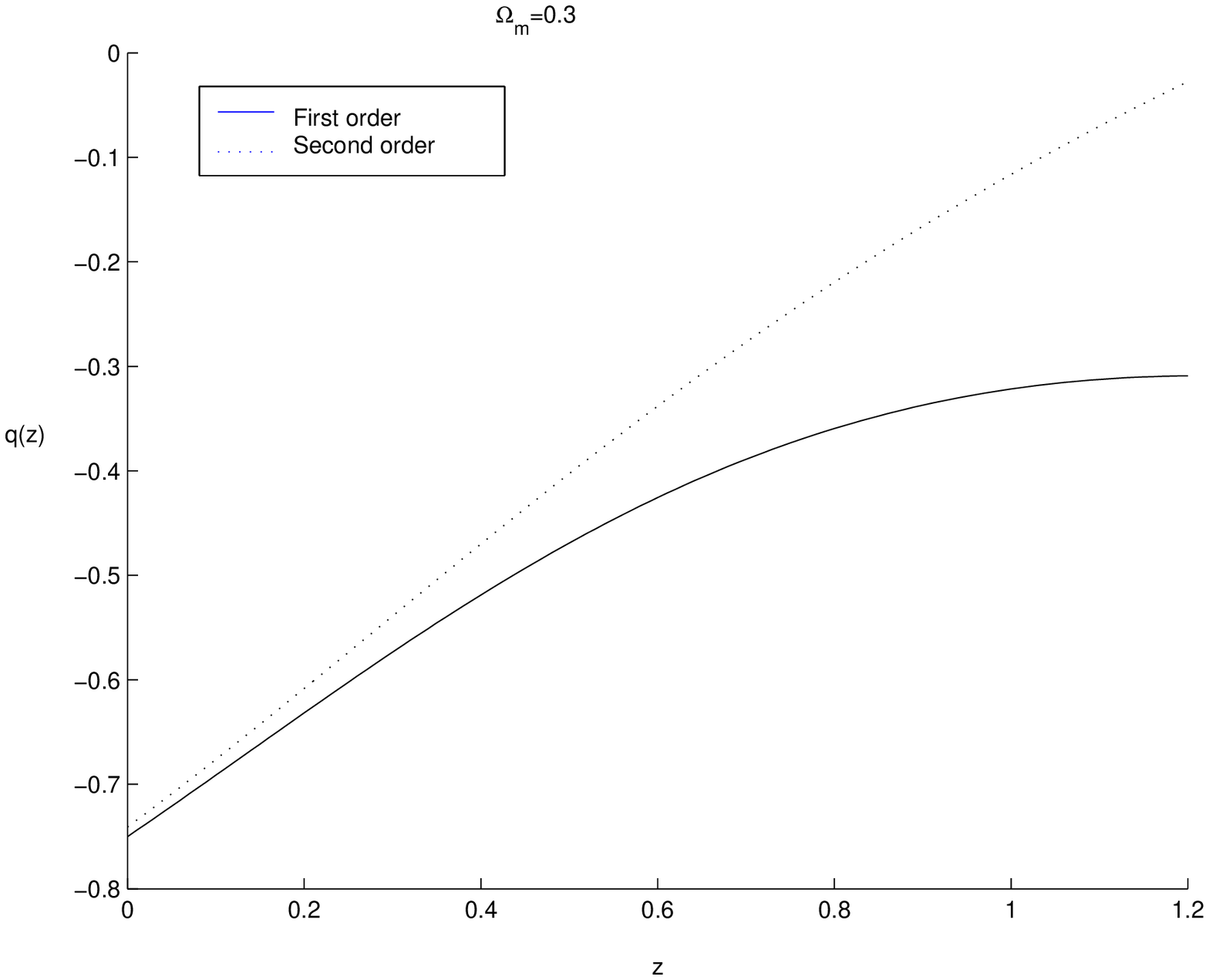}
  \caption{The dependence
of the deceleration parameter $q(z)$ on redshift $z$ for
$\Omega_m=0.1,0.2,0.3$ respectively. The second order effects is
seen to be large.}\label{1}
\end{figure}

Given the observation that our Universe is currently expanding in
an accelerating phase, the deceleration parameter should become
negative in recent cosmological times. The deceleration parameter
is defined by $q=-\ddot{a}/(aH^2)$.

In the case of first order approximation, from eq.(\ref{1stMF}) we
can get
\begin{equation}
q=\frac{\frac{21}{16}\frac{\kappa
\rho}{\alpha}-\frac{99}{32}(\frac{\kappa
\rho}{\alpha})^2-3}{(6+\frac{9}{4}\frac{\kappa
\rho}{\alpha})(\frac{11}{8}\frac{\kappa
\rho}{\alpha}+\frac{1}{2})}\label{1stdec}
\end{equation}

In the case of second order approximation, from eq.(\ref{2ndMF})
we can get
\begin{equation}
q=\frac{\frac{21}{16}\frac{\kappa
\rho}{\alpha}+\frac{33}{32}(\frac{\kappa
\rho}{\alpha})^2+\frac{99}{256}(\frac{\kappa
\rho}{\alpha})^3-3}{(6+\frac{9}{4}\frac{\kappa
\rho}{\alpha})(\frac{11}{8}\frac{\kappa
\rho}{\alpha}+\frac{11}{32}(\frac{\kappa
\rho}{\alpha})^2+\frac{1}{2})}\label{2nddec}
\end{equation}

Since $\kappa \rho/\alpha=(1+z)^3/(1+z_E)^3$, we can get the
dependence of $q(z)$ on redshift, this is shown in Fig.5. It can
be seen that the second order effects can not be neglected for all
the three values of $\Omega_m$.

Combined with the result of Sec.3, we can draw the conclusion that
the second order effects can be neglected on the cosmological
predictions involving only the Hubble parameter itself, e.g. the
various cosmological distances, but the second order effect can
not be neglected in the predictions involving the derivatives of
the Hubble parameter. This later assertion can also be confirmed
by deriving the effective equation of state of this modified
gravity theory.

\textbf{5. The Effective Equation of State}

In this modified gravity theory, following the general framework
of Linder and Jenkins \cite{Linder}, we can define the effective
Equation of State (EOS) of dark energy. Written in this form, it
also has the advantage that many parametrization works that have
been done for EOS of dark energy can be compared with the modified
gravity.

Now following Linder and Jenkins, the additional term in Modified
Friedmann equation really just describes our ignorance concerning
the physical mechanism leading to the observed effect of
acceleration. Let us take a empirical approach, we just write the
Friedmann equation formerly as
\begin{equation}
H^2/H_0^2=\Omega _m(1+z)^3+\delta H^2/H_0^2\label{4.1}
\end{equation}
where we now encapsulate any modification to the standard
Friedmann equation in the last term, regardless of its nature.

Define the effective EOS $\omega _{eff}(z)$ as
\begin{equation}
\omega _{eff}(z)=-1+\frac{1}{3}\frac{d\ln\delta
H^2}{d\ln(1+z)}\label{4.2}
\end{equation}

Then in terms of $\omega _{eff}$, equation eq.(\ref{4.1}) can be
written as
\begin{equation}
H^2/H_0^2=\Omega
_m(1+z)^3+(1-\Omega_m)e^{3\int_0^z{d\ln(1+z')[1+\omega
_{eff}(z')]}}\label{4.3}
\end{equation}

Using the above formalism, we can compute the effective equation
of state corresponding to the first and second order
approximations.

\begin{figure}
  \includegraphics[width=0.4\columnwidth]{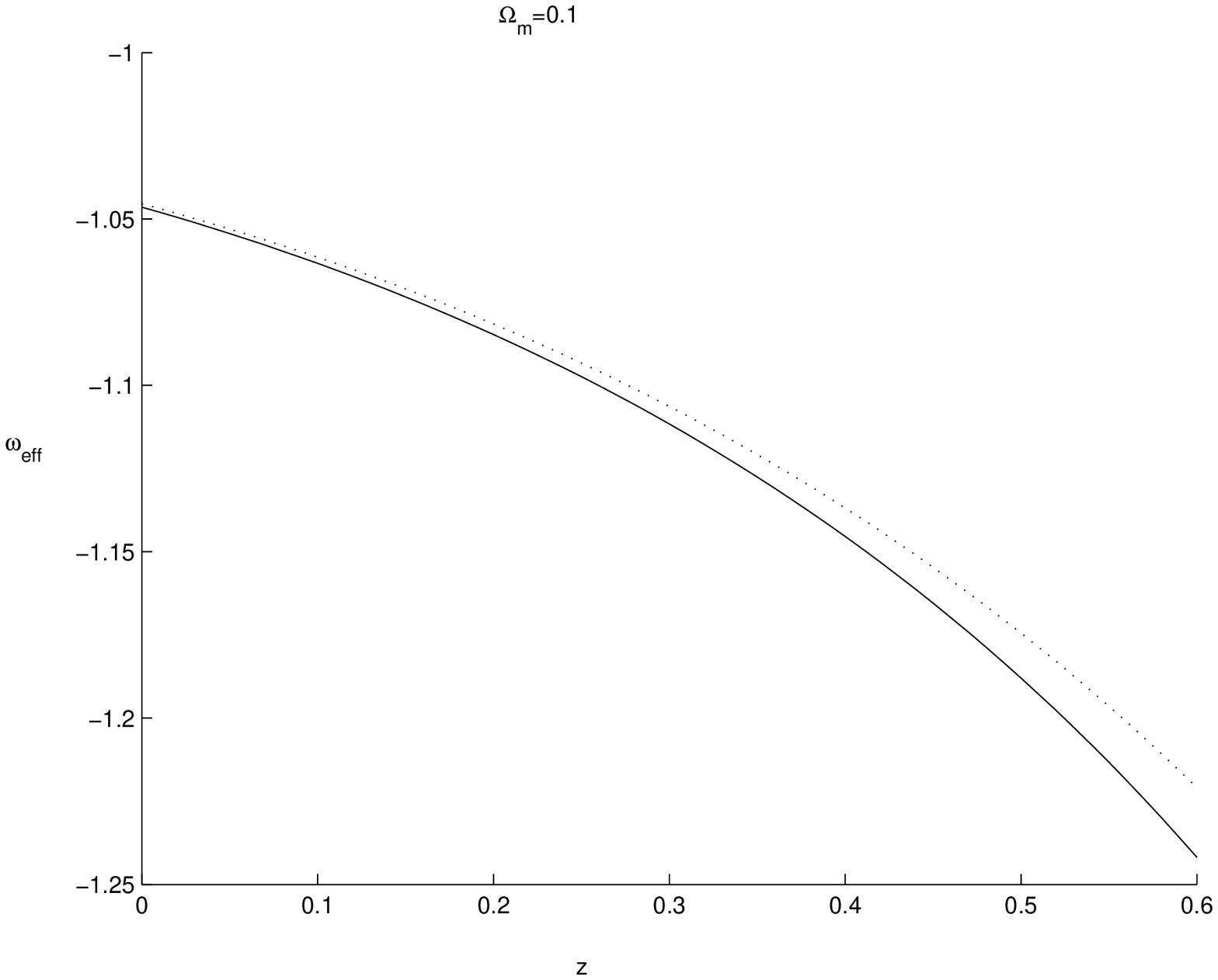}
  \includegraphics[width=0.4\columnwidth]{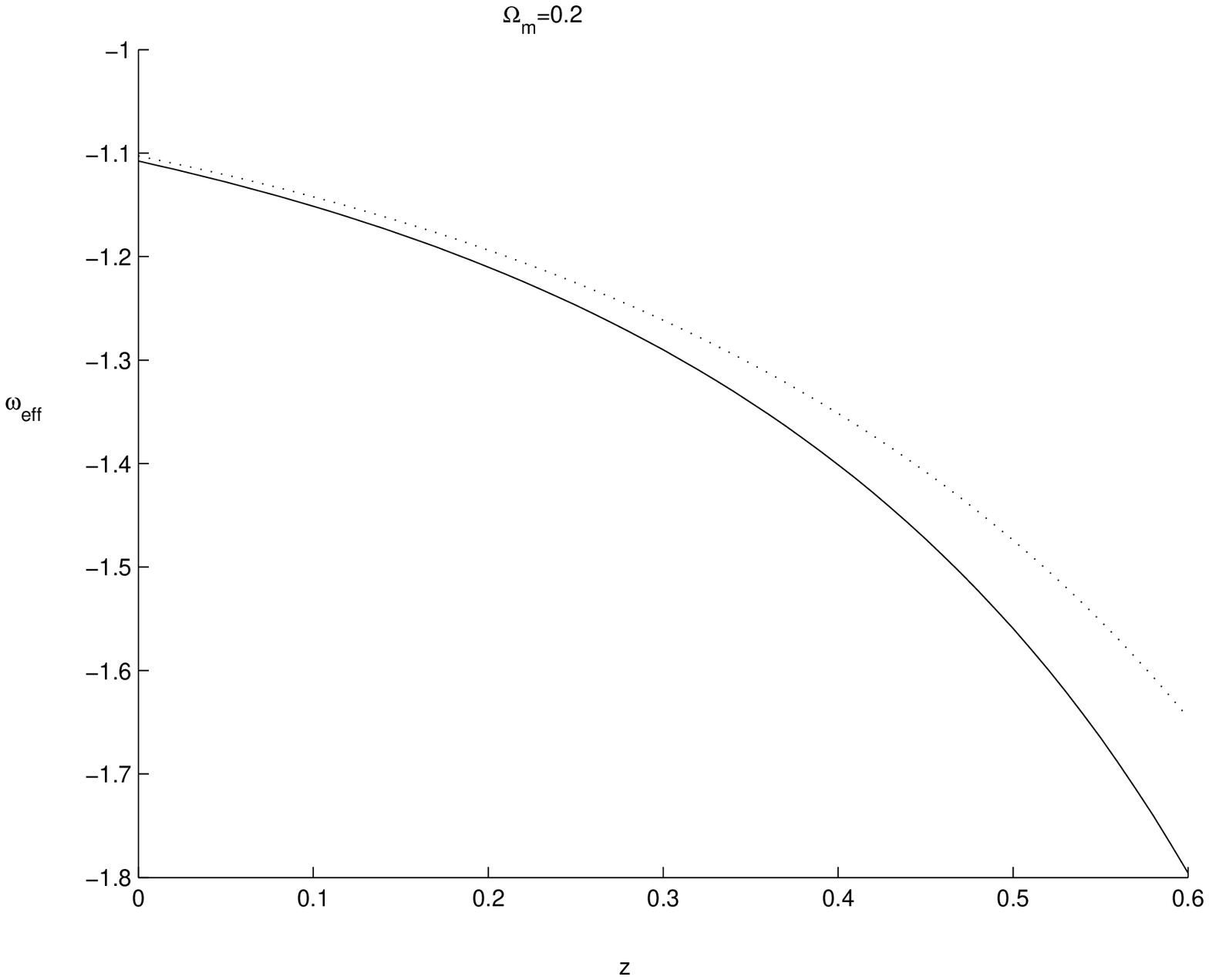}
  \includegraphics[width=0.4\columnwidth]{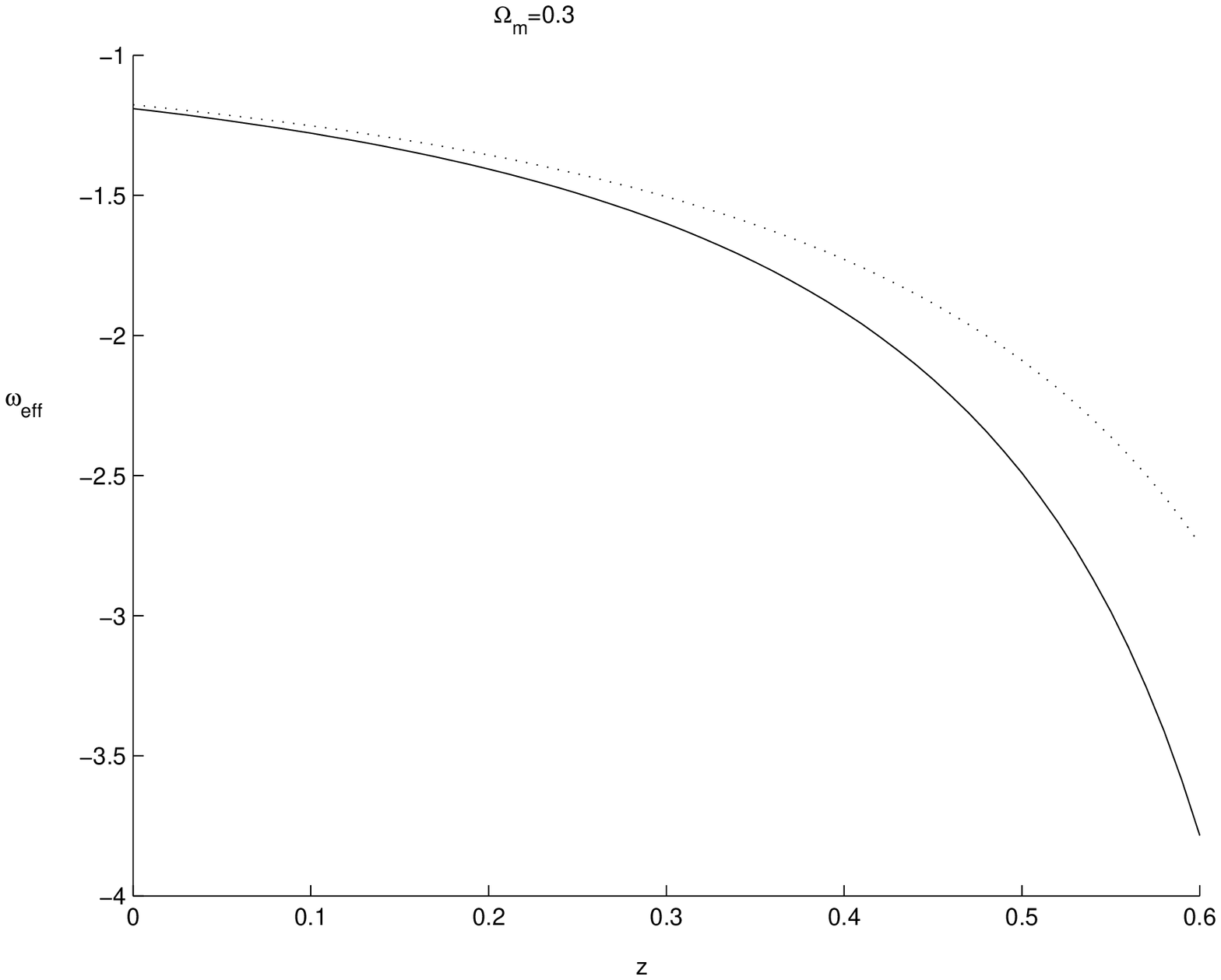}

  \caption{The dependence
of the effective equation of state $\omega_{eff}$ on redshift $z$
in the case of $\Omega_m=0.1, 0.2, 0.3$ respectively. Solid and
dotted lines correspond to first and second order MF equations
respectively. The second order effects is seen to be
large.}\label{1}
\end{figure}

The first order EOS:
\begin{equation}
\omega_{eff}^1=-1-\frac{\frac{9}{16}(\frac{1+z}{1+z_E})^9+3(\frac{1+z}{1+z_E})^6+\frac{13}{8}(\frac{1+z}{1+z_E})^3
}{(-\frac{3}{4}(\frac{1+z}{1+z_E})^6-\frac{5}{8}(\frac{1+z}{1+z_E})^3+\frac{1}{2})(2+\frac{3}{4}(
\frac{1+z}{1+z_E})^3)}\label{1steos}
\end{equation}

The second order EOS:
\begin{equation}
\omega_{eff}^2=-1-\frac{13}{8}\frac{\frac{3}{16}(\frac{1+z}{1+z_E})^9+(\frac{1+z}{1+z_E})^6+(\frac{1+z}{1+z_E})^3
}{(-\frac{13}{32}(\frac{1+z}{1+z_E})^6-\frac{5}{8}(\frac{1+z}{1+z_E})^3+\frac{1}{2})(2+\frac{3}{4}(
\frac{1+z}{1+z_E})^3)}\label{2ndeos}
\end{equation}

These two EOSs are drawn in Fig.6. The difference is quite
obvious. Furthermore, they diverge before $z_E$. Concretely, the
first order EOSs diverge at roughly $z=1.65,1.05,0.75$ for
$\Omega_m=0.1,0.2,0.3$ respectively; the second order EOSs diverge
at roughly $z=1.8,1.15,0.85$ for $\Omega_m=0.1,0.2,0.3$
respectively. In summary, for higher value of $\Omega_m$, the
divergence is more rapid and the first order EOS diverges more
rapid than second order EOS for the same $\Omega_m$, which
indicate that second order EOS is a better approximation to the
full effective EOS.  But we do not know whether the EOS computed
from the full field equations also undergo such a divergence.
Maybe this divergence is only an indication of the breakdown of
perturbation approximation. Note $w_{eff}<1$, which is also a
possible behavior of EOS, see ref.\cite{Carroll2}.

\textbf{6. Discussions and Conclusions.}

Applying Palatini formalism to the modified Einstein-Hilbert
action is an interesting approach that is worth further
investigations. In particular, it is free of the sort of
instability encountered in the metric variation formalism as
argued by Dolgov et. al. \cite{Dolgov}. And more importantly, we
have showed that the MF equations fit the SNe Ia data at an
acceptable level. In order to compute other cosmological
parameters such as the age of the universe, we need the full MF
equation, which has been investigated by us elsewhere
\cite{Wang3}. It is shown there that the age of the universe is
compatible with the age of the globular agglomerates observed
today.

We have discussed the effects of second order approximation. We
showed that the second order effects can be neglected on the
cosmological predictions involved only the Hubble parameter
itself, e.g. the various cosmological distances, but the second
order effects are large in the predictions involved the
derivatives of the Hubble parameter. Thus, in the future work
dedicated to discuss the cosmological effects or local gravity
effects of this modified gravity theory, second order MF equation
(\ref{2ndMF}) seems a better starting point. Of course, whether
the effects of third order effects or even the non-perturbative
effects can not be negligible in redshift smaller than $z_E$ for
some situations deserve further investigations.

In summary, adding a $R^{-1}$ or the like terms to the
Einstein-Hilbert action is an interesting idea, which may
originate from some Sring/M-theory \cite{jap}, and looks like a
possible candidate for the explanation of recent cosmic expansion
acceleration fact. We can see such modifications may accommodate
the update observational data indicating our Universe expansion is
accelerating without introducing the mysterious so called Dark
Energy. However, so far as we know the update experimental
testings to General Relativity Gravity theory within the Solar
system \cite{data} have repeatedly confirmed that the GR is right;
now for larger cosmic scale if we should modify the conventional
gravity theory to confront cosmological observations, it is vital
to both our basic understanding of the Universe and  development
for fundamental physical theories.

\textbf{Acknowledgements}

We would like to thank the referees' very helpful comments which
have improved this paper greatly.
 P.W. wishes to thank Professor A. Lue
and R. Scoccimarro for valuable discussions. X.H.M. has benefitted
a lot by helpful discussions with R.Branderberger, D.Lyth,
A.Mazumdar, L.Ryder, X.P.Wu, K.Yamamoto and Y. Zhang.  This work
is partly supported by China NSF, Doctor Foundation of National
Education Ministry and ICSC-world lab. scholarship.

\begin{appendix}
\end{appendix}

\end{document}